\documentclass{article}
\usepackage[dblblindworkshop, final]{neurips_2025}

\usepackage[utf8]{inputenc} 
\usepackage[T1]{fontenc}    
\usepackage{hyperref}       
\usepackage{url}            
\usepackage{booktabs}       
\usepackage{amsfonts}       
\usepackage{nicefrac}       
\usepackage{microtype}      
\usepackage{xcolor}         
 \usepackage{graphicx}

\title{Semantically Annotated Multimodal Dataset for \\ RF Interpretation and Prediction}

\author{%
  S. Blandino$^{1,2}$, J. Senic$^1$, R. Caromi$^1$, S. Berwerger$^1$, A. Bodi$^{1,2}$, C. Gentile$^1$, N. Golmie$^1$ \\
  $^1$National Institute of Standards and Technology (NIST) \\
  $^2$Prometheus Computing LLC\\
  \texttt{steve.blandino@nist.gov} \\
}

\begin{document}

\vspace{-1cm}
\maketitle

\vspace{-1cm}
\section{Introduction: dataset rationale}

Imagine being tasked with deploying a wireless or cellular network that must provide full coverage in an environment filled with physical barriers such as walls, buildings, and heterogeneous terrain.
In such complex settings, incomplete knowledge of material properties and environmental features renders electromagnetic simulations both computationally intensive and susceptible to inaccuracies. 
In practice, the most reliable solutions come from labor-intensive field measurements with specialized instrumentation, which provide the empirical insight needed to understand how radio-frequency (RF) signals propagate in these complex environments [1].
Breakthrough advances in RF propagation prediction - particularly those that go beyond incremental improvements - could revolutionize wireless system design and significantly speed up the rollout of new technologies.
 
Current limitations in wireless modeling and RF-based AI are driven primarily by the lack of high-quality, measurement-based datasets. Acquiring high-resolution radio-frequency (RF) measurements to characterize how radio signals propagate and scatter in an environment is an inherently challenging task, typically demanding complex and costly hardware setups. 
Measurement-based datasets usually take the form of RF heatmaps, which are high-dimensional data representations akin to 3D images, but instead of color values, each "voxel" encodes signal intensity distributed across angle, delay, and time. While visually analogous to images or volumetric medical scans, RF heatmaps present a unique interpretability challenge that is unique to their domain: they lack direct geometric or semantic context. 
Without explicit labels or visual references, it is difficult to identify which part of the measured RF signals corresponds to specific physical objects, surfaces, or environmental interactions. 
This interpretability gap significantly constrains the development of robust, data-driven machine learning (ML) models, as supervised learning depends critically on reliable ground-truth associations between the data and labels.
\section{Dataset}

To address this significant bottleneck, we envision the creation of a new class of multimodal datasets specifically designed to bridge the gap between RF signals and their underlying physical causes. Instead of restricting RF interpretation to a single sensor, such datasets would \textbf{combine RF measurements with diverse auxiliary modalities}, each capturing a different facet of the environment, its geometry, motion, material properties, or dynamic processes. These modalities could include high-resolution imagery from conventional or hyperspectral cameras, dense 3D reconstructions from lidar, acoustic recordings that capture structural vibrations, radar or ultrasonic scans at complementary frequencies, or any other sensing technology that enriches the physical description of the scene.

To achieve this vision, our proposed data collection will span {a wide range of environments and scenarios}. We plan to capture data (RF, images and lidar point clouds) in diverse settings, from controlled indoor laboratories and cluttered rooms to complex outdoor environments. The scenarios will feature a variety of static and dynamic objects, including autonomous robots navigating complex paths and human subjects engaged in a spectrum of activities. These activities will range from large-scale movements like walking  to more subtle actions such as hand gestures, sitting, and even the micro-movements associated with sleeping such as chest motion induced by hearth and breathing activities. This comprehensive approach will ensure the resulting dataset {captures a rich and varied distribution of RF phenomena}, providing a robust foundation for developing generalizable predictive AI models.
A major technical challenge lies in achieving precise temporal and spatial co-registration with RF measurements. This hurdle could be overcome by using context-aware channel sounders [2], which provides the robust spatial alignment and temporal synchronization necessary to ensure every recorded RF event can be unambiguously associated with a specific physical structure. In addition, digital replica reconstructions will provide precise voxel-level annotation by linking every element of the RF signal to a physical interpretation. The resulting dataset is not only complex-valued and multimodal but also enriched with semantic labels that transform raw propagation measurements into interpretable and scientifically useful data.

With this critical infrastructure established, the primary remaining bottleneck is the resource-intensive nature of large-scale data acquisition. Addressing this challenge requires not only efficient measurement procedures, but also a sustainable governance and dissemination framework that enables coordinated data collection, reuse, and long-term maintenance. Accordingly, this dataset and its associated methodology are contributed within the framework of the \textbf{NextG Channel Model Alliance} hosted at nextg.nist.gov, a NIST-led open research initiative that provides governance, continuity, and best practices for channel modeling assets across the standards lifecycle. The Alliance supports transparent versioning, curated public releases, and community-driven refinement of datasets, facilitating scalable acquisition efforts and broad adoption across the research and standards communities.

\section{AI task definition and acceleration potential}
The dataset will enable transformative advances across several AI-driven scientific questions.
At its core lies a grand challenge: \textit{Can AI learn a universal mapping between scene perception and RF propagation?}. This challenge motivates both forward and inverse learning tasks. 

The dataset supports a forward prediction task where an AI model learns to \textbf{predict a complex RF heatmap from corresponding perception data} (e.g., camera and lidar). Once such a model is developed, it could directly infer RF propagation characteristics for a new environment using only its visual and geometric data as input. This would revolutionize wireless system design by enabling rapid and accurate simulations without the need for labor-intensive physical RF measurements.
Equally significant is the inverse task: \textbf{inferring scene geometry and semantics directly from complex RF signals}. This inverse semantic segmentation of RF data represents a fundamental leap toward a form of RF-based perception analogous to vision, laying the foundation for entirely new RF-based sensing capabilities.
Moreover, the dataset enables \textbf{generative modeling tasks}, where models synthesize physically consistent RF scenarios  creating realistic environments that would otherwise demand prohibitively costly measurements and augmenting existing datasets with diverse, scalable RF conditions. Such capabilities naturally extend to digital twins, where AI-generated RF environments can populate virtual replicas of real-world systems for design, testing, and automation.

The release of this dataset has the potential to accelerate discovery across multiple domains. For wireless system design, {physics-informed ML models}, such as differentiable ray-tracers [3], to be trained directly on measured data rather than simplified simulations, will dramatically reduce the gap between simulations and reality. This would facilitate rapid design loops for wireless system optimization, reducing simulation times from hours (full-wave electromagnetic simulations) or minutes (ray-tracing simulations) down to milliseconds, thereby supporting advanced optimization of wireless systems.
Moreover, by providing a realistic, complex-valued, multimodal dataset, we anticipate {substantial cross-disciplinary} spillover. The dataset will attract researchers from computer vision, robotics, and applied physics communities, {promoting algorithmic breakthroughs} in complex-domain ML, multimodal sensor fusion, and physics-informed neural network designs. Finally, {the dataset will serve as a robust benchmark}, driving innovation through the development of new loss functions and training strategies that blend  electromagnetic theory with modern ML methods.

In summary, this semantically segmented multimodal RF dataset will profoundly transform downstream science, unlocking new capabilities in wireless perception, autonomous navigation, extended reality applications, and ultimately advancing both theoretical understanding and practical implementations of physics-aware artificial intelligence.


\newpage
\appendix

\section{Technical Appendices and Supplementary Material}

To verify the feasibility of the envisioned dataset, we captured an initial dataset using a specific multimodal configuration combining synchronized RF heatmaps, lidar-derived point clouds, and high-resolution panoramic camera imagery.
This appendix provides the technical background supporting the feasibility study described in the main submission.
While the core proposal focuses on the dataset vision, and its scientific impact, the following sections detail the hardware platform, sensing configuration, and annotation pipeline used to produce the initial dataset.
These details are not required to understand the broader proposal but are included here to demonstrate that the approach is technically viable and to document the specific system used for the proof-of-concept capture.
The methods described are general and can be adapted to other RF sensing platforms or multimodal measurement setups.

\subsection{RF Measurement System}

\begin{figure}
  \centering
  \includegraphics[width=0.5\columnwidth]{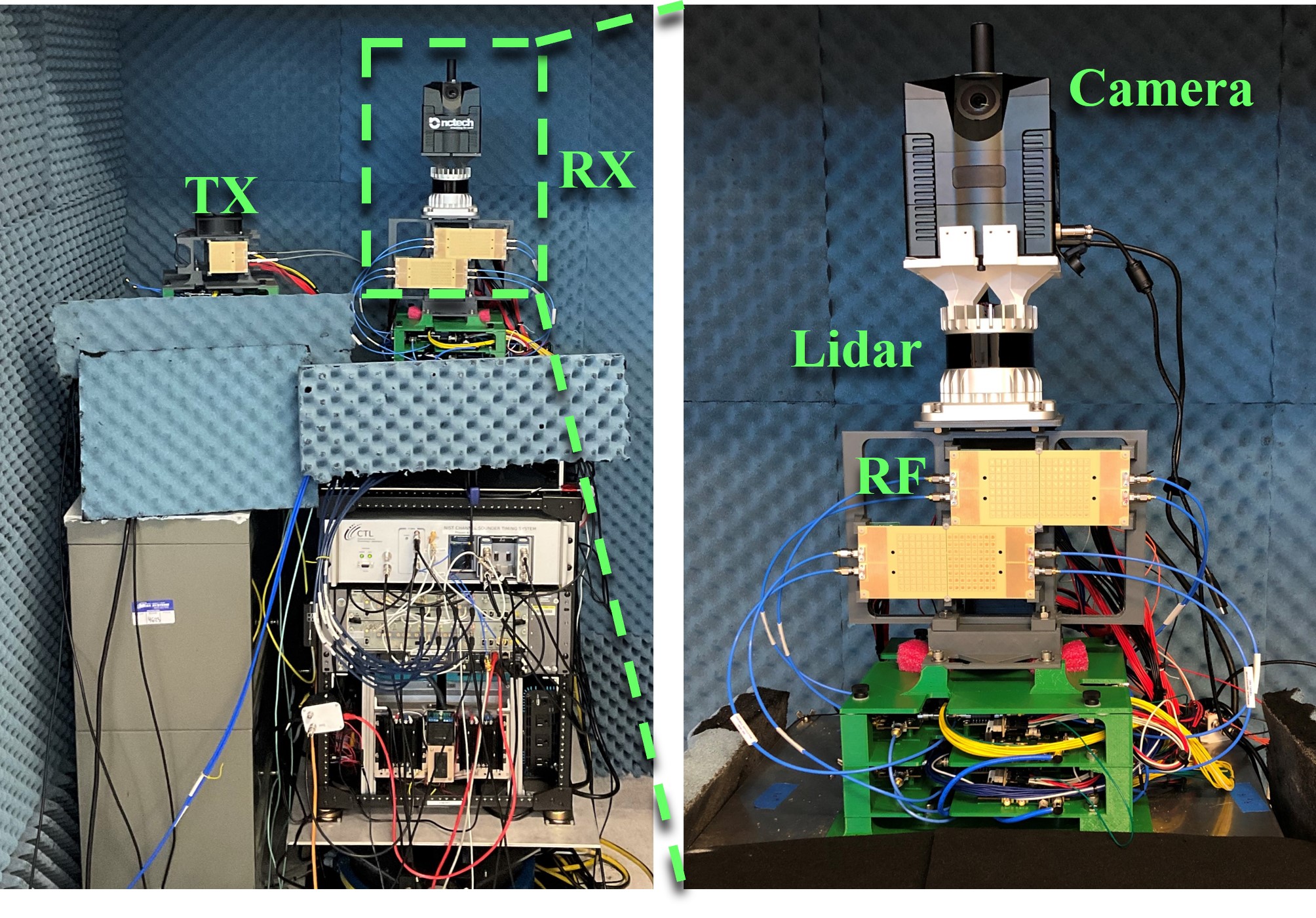}\vspace{-2mm}
  \caption{28\,GHz context-aware channel sounder.}
  \label{fig:sounder}
\end{figure}

The feasibility study was conducted using a 28.5\,GHz context-aware channel sounder, shown in Figure~1, equipped with an \(8{\times}8\)  phased-array transmitter and four \(8{\times}8\)  phased-array receivers stacked to form a 256-element array.
The transmitter used analog beamforming to achieve quasi-omnidirectional coverage and transmitted a 2047-chip PN sequence with 2\,GHz bandwidth at 30\,dBm EIRP. Phase coherence between TX and RX was maintained via a Rubidium clock with optical trigger distribution.

Digital beamforming synthesized 4500 digital beams across a 
\(90^\circ \times 50^\circ\) FoV with \(1^\circ\) steering steps. Each RF frame (acquired every 2.6\,ms) provided a complex-valued 4D heatmap over azimuth, elevation, delay, and time with 0.0625\,ns delay resolution.

\subsection{Camera and Lidar Measurement System}
A mechanical lidar scanner (2048 azimuth × 128 elevation points per frame, 1 cm range resolution) and a panoramic RGB camera (11 000 × 5 500 pixels) were co-mounted with the RF array at the RX. Lidar operated at 100\,ms per frame and the camera at 140 ms per frame. All systems were synchronized to the Rubidium clock; spatial registration was refined using metallic markers visible in all modalities.

\subsection{Digital Twin Registration}

High-resolution watertight meshes of human subjects were generated from RGB panoramas using the ECON algorithm [4], which reconstructs full-body geometry including clothing details. These canonical meshes were registered to the lidar point cloud using 3D keypoints derived from HRNet detections in both image and lidar domains [5].
A rigid similarity transform (scale, rotation, translation) aligned the canonical mesh to the measurement frame for each time step, producing a registered digital twin that followed the subject’s posture and motion.

\subsection{Voxel-Level Annotation of RF Heatmaps}

\begin{figure*}[!h]
    \centering

    \begin{minipage}[b]{0.46\textwidth}
        \centering
        \includegraphics[width=\textwidth]{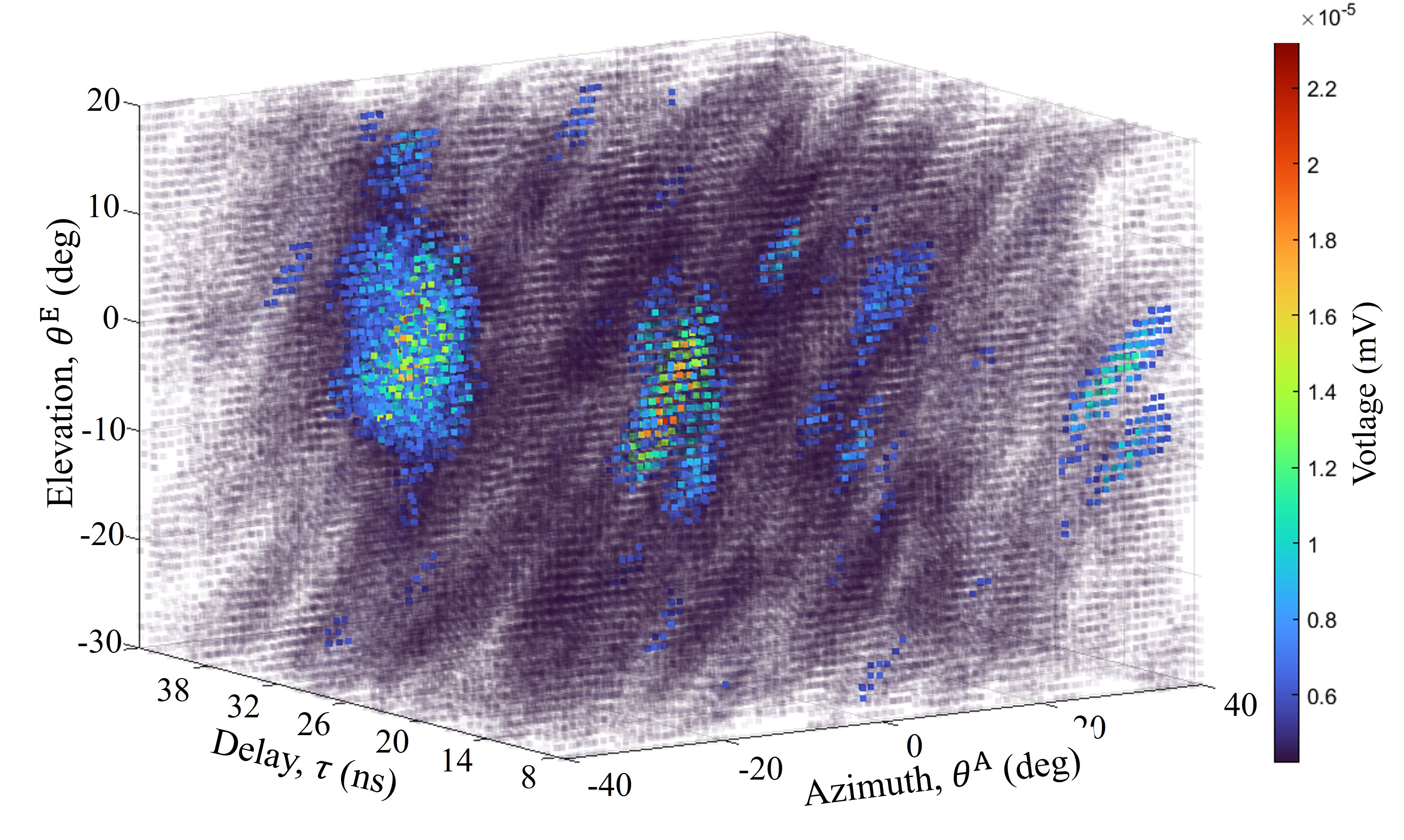} \\
        \small (a) Generated heatmap at $t{=}$0.086\,s
    \end{minipage}
    \hspace{0.05\textwidth}
    \begin{minipage}[b]{0.39\textwidth}
        \centering
        \includegraphics[width=\textwidth]{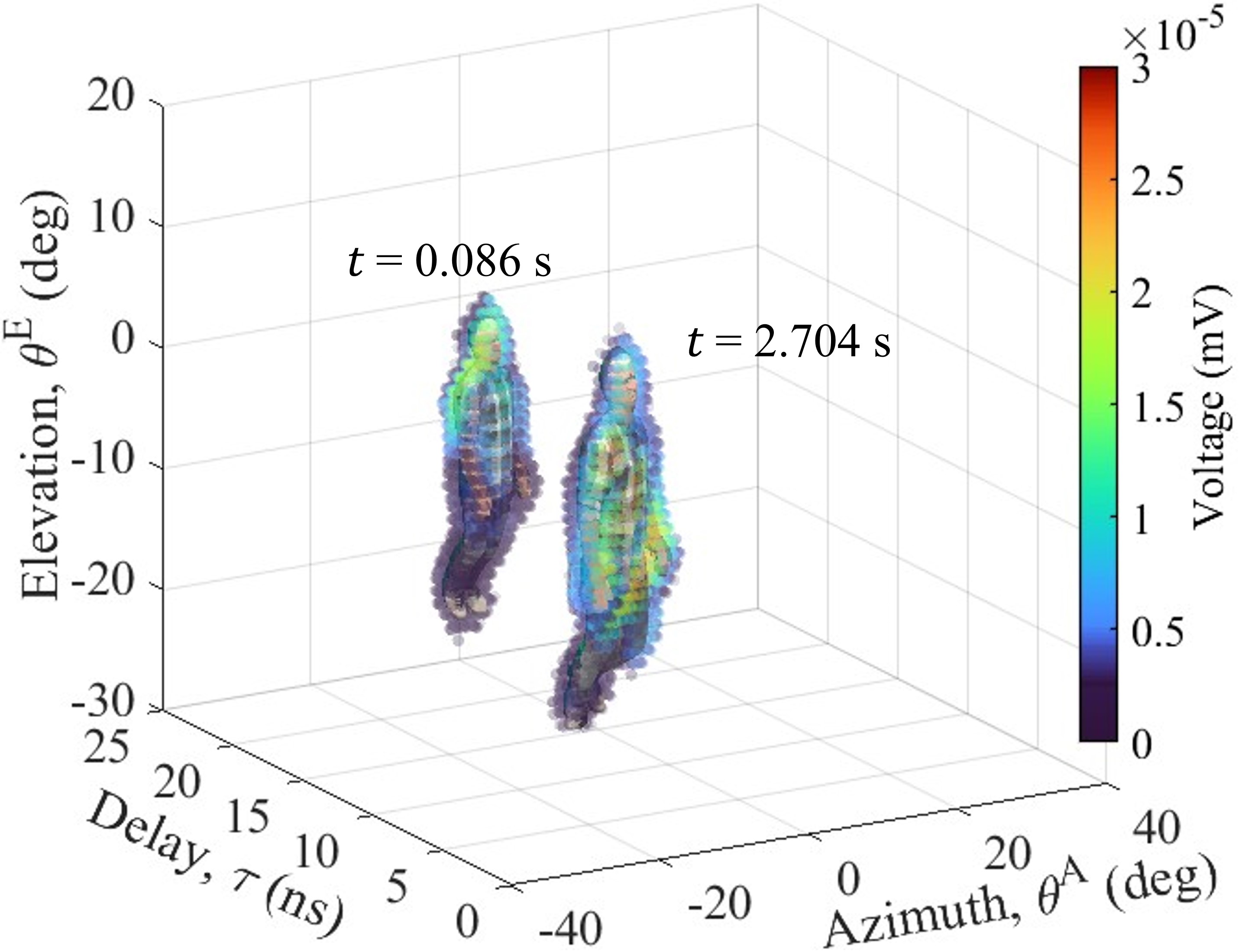} \\
        \small (b) Twin and segmented heatmap overlay at two times
    \end{minipage}

    \vspace{3mm}

    \begin{minipage}[b]{0.46\textwidth}
        \centering
        \includegraphics[width=\textwidth]{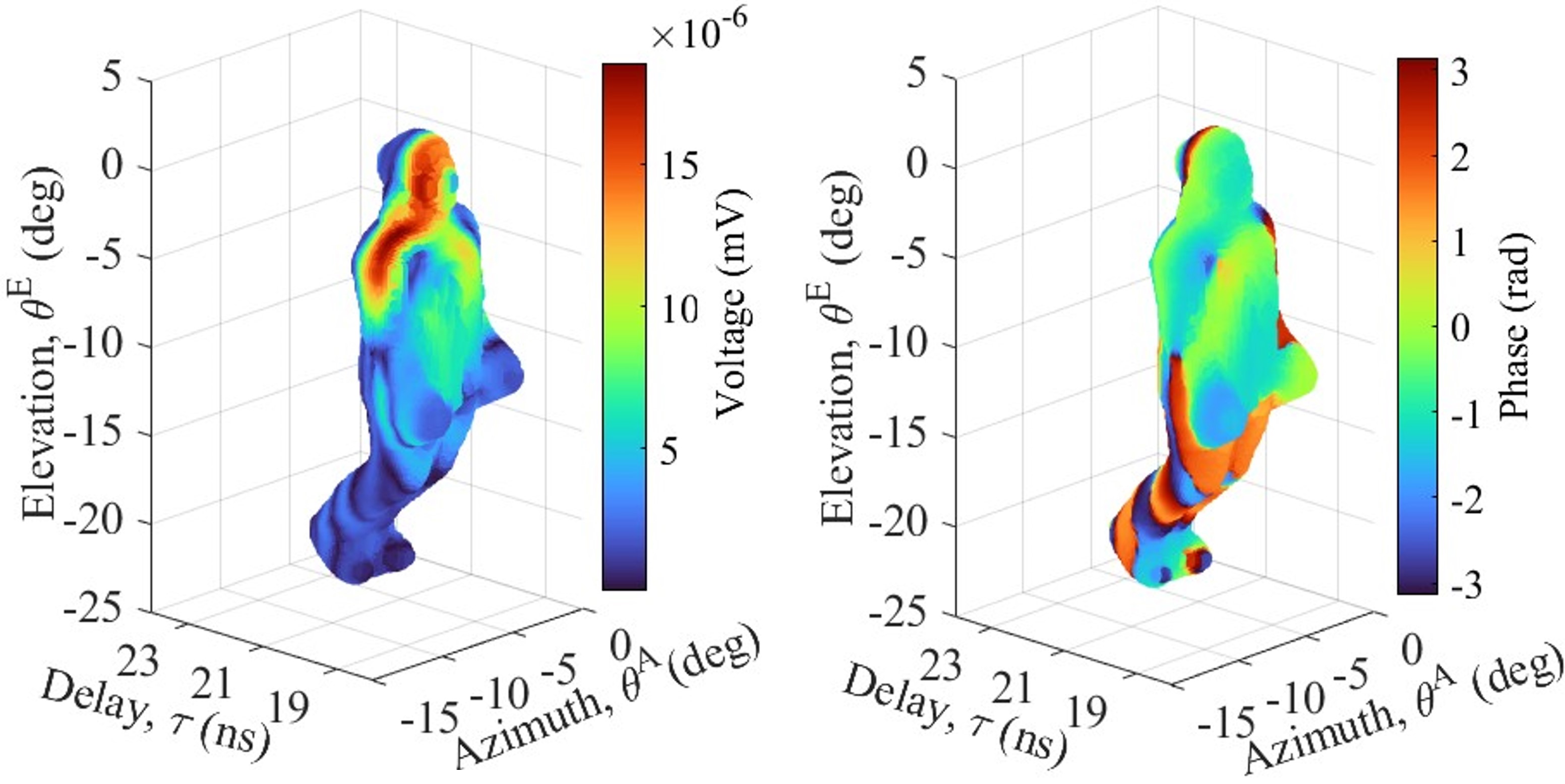} \\
        \small (c) Magnitude and phase of segmented heatmap at $t{=}$0.086\,s
    \end{minipage}
    \hspace{0.05\textwidth}
    \begin{minipage}[b]{0.46\textwidth}
        \centering
        \includegraphics[width=\textwidth]{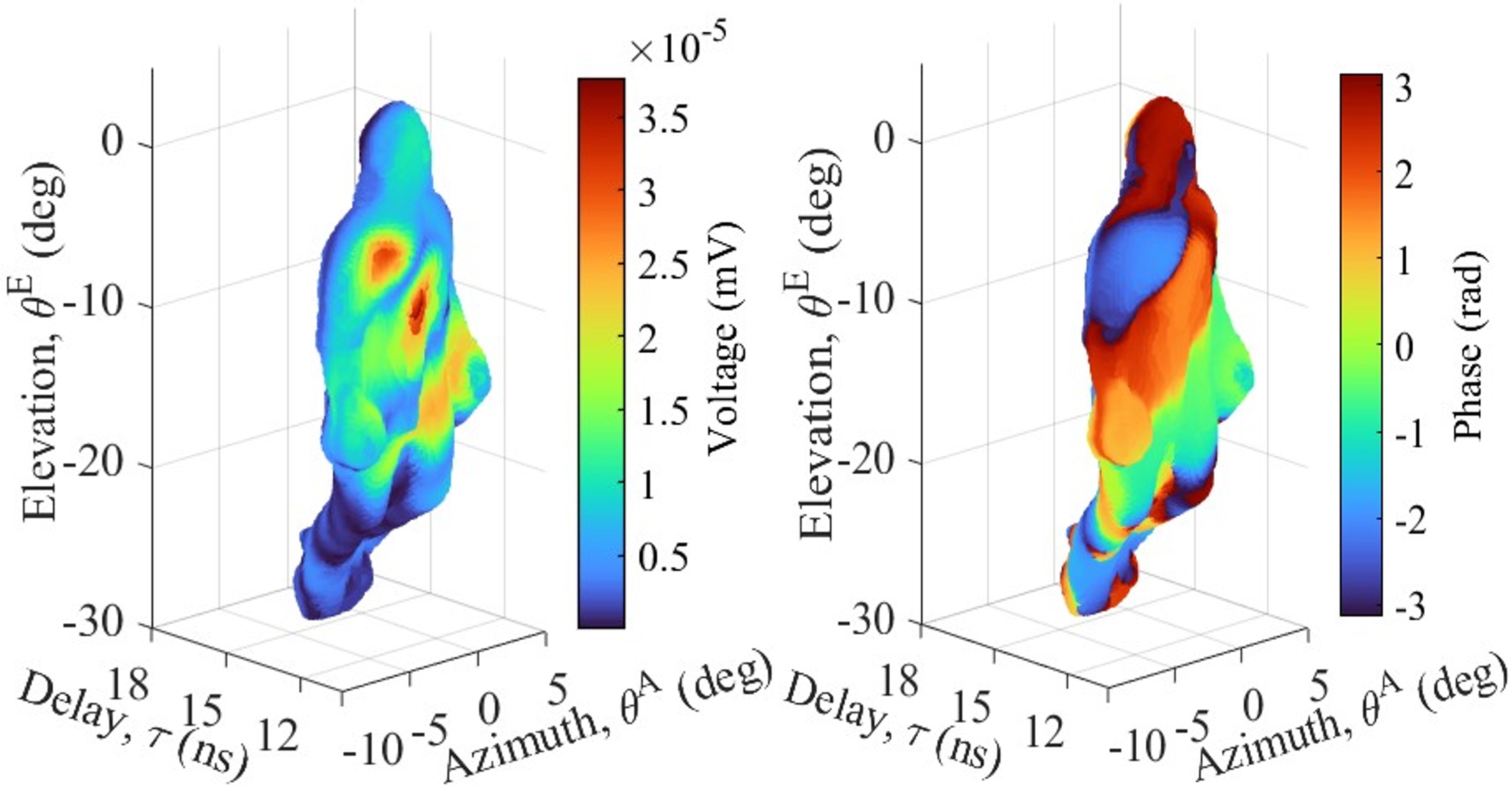} \\
        \small (d) Magnitude and phase of segmented heatmap at $t{=}$2.704\,s
    \end{minipage}

    \caption{Visualization of RF heatmap segmentation using a registered digital twin.}
    \label{fig:heatmap_segmentation_visual}
\end{figure*}
\normalsize

The registered digital twin was projected into the spherical-delay coordinate system of the RF heatmap using bistatic geometry. Each RF voxel was labeled according to whether it intersected the twin surface, enabling direct semantic association between measured RF signals and the object and body part that produced them.
This geometry-based segmentation does not require motion assumptions and preserves both specular and diffuse scattering.

Fig.\ref{fig:heatmap_segmentation_visual} illustrates the segmentation process. Subfig.(a) shows the full RF heatmap at $t$ = 0.086,s, rendered with semi-transparent opacity proportional to voxel magnitude to reveal internal structure.
Before applying the segmentation and annotation framework, the RF heatmap—despite its high resolution—offered no direct way to associate individual signal components with specific physical objects or body parts in the scene, making it largely uninterpretable from a semantic standpoint.
Subfig.(b) overlays segmented heatmaps at two non-overlapping time snapshots: $t$ = 0.086\,s and $t$ = 2.704\,s, from a male subject walking linearly over 5.2\,s. Transparency reveals the underlying digital twin in each frame.
Subfigs (c) and (d) zoom in on the two segmented frames in (b), using complex-valued RF data to show both magnitude and phase. 
Through this process, each magnitude and phase value in the heatmap is annotated with the corresponding coordinate obtained from the registered digital replica, establishing a direct link between RF measurements and their precise physical origin.

\subsection{Initial proof-of-concept}

The initial proof-of-concept capture was carried out across nine distinct RF antenna positions and orientations, with three human subjects performing controlled walking sequences. Informed consent was obtained from individual participants included in the study. In total, the acquisition produced approximately 50 000 annotated RF heatmaps, 950 panoramic RGB frames, and 1 400 lidar point clouds. Each RF frame was temporally and spatially aligned with its corresponding visual and geometric measurements, allowing voxel-level semantic annotations to be generated with high geometric accuracy. This dataset confirmed that the proposed approach can reliably achieve precise co-registration and annotation in practice, providing a solid foundation for scaling to additional sensing modalities and more complex environments.

\clearpage

\subsection*{REFERENCES}

{
\small

[1] A. Alkhateeb et al., "DeepSense 6G: A Large-Scale Real-World Multi-Modal Sensing and Communication Dataset," in I\textit{EEE Communications Magazine}, vol. 61, no. 9, pp. 122-128, September 2023, doi: 10.1109/MCOM.006.2200730.

[2] A. Bodi, S. Berweger, R. Caromi, J. Bang, J. Senic and C. Gentile, "AI-Based Environment Segmentation Using a Context-Aware Channel Sounder," \textit{2024 18th European Conference on Antennas and Propagation (EuCAP)}

[3] J. Hoydis et al., "Learning Radio Environments by Differentiable Ray Tracing," in \textit{IEEE Transactions on Machine Learning in Communications and Networking}, vol. 2, pp. 1527-1539, 2024, doi: 10.1109/TMLCN.2024.3474639. 

[4] Y. Xiu, J. Yang, X. Cao, D. Tzionas and M. J. Black, "ECON: Explicit Clothed humans Optimized via Normal integration," \textit{2023 IEEE/CVF Conference on Computer Vision and Pattern Recognition (CVPR)}

[5] J. Wang et al., "Deep High-Resolution Representation Learning for Visual Recognition," in IEEE Transactions on Pattern Analysis and Machine Intelligence, vol. 43, no. 10, pp. 3349-3364, 1 Oct. 2021, doi: 10.1109/TPAMI.2020.2983686

}

\end{document}